\documentclass{elsart}

\usepackage{psfig}

\usepackage{natbib}

\begin{document}

\runauthor{Pfeiffer, Appenzeller and Wagner}


\begin{frontmatter}

\title{Coronal lines and the warm X-ray absorber in Seyfert~1 Galaxies}

\author[PSU]{M. Pfeiffer}
\author[PSU]{I. Appenzeller}
\author[PSU]{S. Wagner}
\address[PSU]{Landessternwarte Heidelberg, Koenigstuhl, 69117 Heidelberg}

\begin{abstract} 
The connection between the coronal lines and the warm absorber is examined systematically. 
In an earlier work it was found that the coronal line emitting plasma and the warm absorber gas 
share the same density and temperature. 
If there is a connection between the warm absorber gas and the 
forbidden high-ionization line (FHIL) plasma, one can use the profiles 
of coronal lines to derive the kinematics and dynamics of the warm absorber due to the 
high spectral resolution available in the optical range. Further support for a connection is 
a correlation between the equivalent width of [Fe X] 6375~\AA\ and the 
ROSAT spectral index found for an optically selected sample. 
For X-ray selected objects with absorption edges observed by ASCA, we looked for a correlation 
between the coronal lines and the warm absorber gas. A direct correlation cannot be confirmed. 
\end{abstract}


\end{frontmatter}


\section{Introduction}

Coronal lines are used as a tool to investigate the intermediate region between the 
NLR and BLR 
(e.g.\ Osterbrock 1981; Appenzeller \& \"Ostreicher 1988; 
Giannuzzo et al.\ 1990). 
Coronal lines are highly ionized emission lines originating from forbidden transitions with an 
ionization potential of at least 90 eV. Coronal lines are therefore also called forbidden 
high-ionization lines, abbreviated as FHILs. The existence of forbidden lines indicates that in the respective 
region the densitites are low enough that the forbidden transitions are not suppressed by collisions. 
The critical densitites are of the order of $10^7$--$10^{10}$~cm$^{-3}$. Due to the high ionization potential, 
coronal lines are an indication of highly energetic 
processes. The density and temperature of the coronal line region 
are of the order of 10$^6$ cm$^{-3}$ and 10$^5$ K (Erkens, Appenzeller \& Wagner 1997). 
These conditions are similiar to those of the 
partially ionized gas producing the O~VII and O~VIII absorption edges 
observed with ASCA, called the warm absorber 
(e.g.\ Nandra \& Pounds 1992; Reynolds \& Fabian 1995). 
The warm absorber is presumably located between the BLR and 
the NLR, and it is found in approximately half of Seyfert~1 galaxies 
(Reynolds 1997). Coronal lines and X-ray 
absorption edges may form in the same plasma. Further support for a connection 
between the warm absorber and the FHILs is given by a correlation between 
the ROSAT PSPC spectral index and the equivalent 
width of [Fe X] 6375 \AA \ found by Erkens, Appenzeller \& Wagner (1997) for an optically selected sample 
(see Figure~1). [Fe X] 6375 \AA \ is a good probe of the FHIL plasma because, as a rule, 
\begin{figure}[htb]
\centerline{\psfig{figure=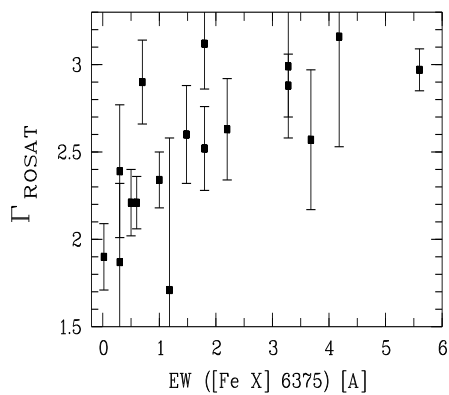,height=2.2truein,width=2.5truein,angle=-90}}
\caption{Correlation between the equivalent width of [Fe X] 6375 \AA \ and the 
ROSAT PSPC spectral index for the optically selected sample.}
\end{figure}
this line is relatively strong. Furthermore, lower ionization 
lines such as [Fe VII] can be stimulated thermally as well. 
If there is a connection between the warm absorber gas and the 
FHIL plasma, one can use line profiles of FHILs to derive the 
kinematics and dynamics of the warm absorber due to the 
higher spectral resolution in the optical than in the X-ray range. 
A connection may also answer the question of whether the same ionization 
mechanism drives FHILs and the warm absorber.

To investigate this question, we observed an X-ray 
selected sample of 16 Seyfert galaxies in the optical range. 
Line properties of a further 4 objects were taken from the literature.  
The selection criterion was that O VII and O VIII absorption edges were observed with ASCA.   
Nearly half of the objects are NLS1s (10 out of 21), the 
other half are Seyfert~1s, and one object is a Seyfert~2.
Though NLS1s are defined by optical criteria, our 
sample was X-ray selected. Then, why are so many NLS1s in our sample? 
Optically selected Seyfert~1 samples contain approximately 10\% NLS1s (Stephens 1989);
this percentage rises to 16--50\% 
when the sample is selected by soft X-ray criteria (Puchnarewicz et al.\ 1992). 
So, the percentage of 50\% is at the upper limit given for soft X-ray selected samples. 
To check how significant the considered correlation is, we calculated correlation coefficients, 
specifically the value of Kendall's $\tau$. The Kendall 
$\tau$ value of the correlation between the ROSAT PSPC spectral index and the 
equivalent width of [Fe X] 6375 \AA \ for the objects of the 
optically selected sample of Erkens, Appenzeller \& Wagner (1997)
is 0.55, corresponding to a probability of 0.09\% that the distribution is only by chance. 
The correponding values for the X-ray selected sample (see Figure~2) 
are 0.21 and a probability of 11\%.
\begin{figure}[htb]
\centerline{\psfig{figure=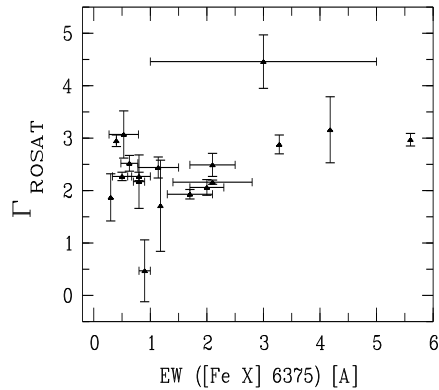,height=2.2truein,width=2.5truein,angle=-90}}
\caption{Correlation between the equivalent width of [Fe X] 6375 \AA \ and the 
ROSAT PSPC spectral index for the X-ray selected sample.}
\end{figure}
With the spectral index alone, one cannot determine if the
correlations shown in Figures~1 and 2 really demonstrate
a connection between the warm absorber and the coronal lines
(ROSAT spectra can appear steep due to the presence of a warm
absorber, but they can also be intrinsically steep). 
The optical depths of the O~VII and the O~VIII edges give 
direct information about the warm absorber. They correlate with 
each other, as shown in Figure~3.
\begin{figure}[htb]
\centerline{\psfig{figure=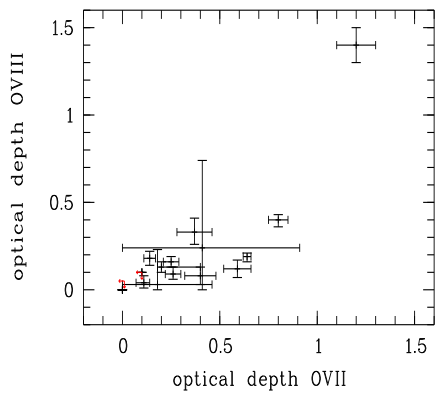,height=2.2truein,width=2.5truein,angle=-90}}
\caption{The optical depths of the O~VII and O~VIII absorption edges.}
\end{figure}
For the optically selected sample alone a comparison between equivalent width and optical depth 
is not possible since ASCA data are available for only three objects of this sample. 
In Figure~4, the optical depth of the O~VII edge is plotted against the equivalent width of the 
[Fe X] 6375~\AA\ line. No correlation can be seen. 
\begin{figure}[htb]
\centerline{\psfig{figure=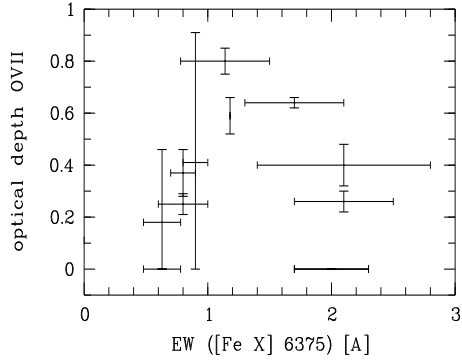,height=2.0truein,width=2.5truein,angle=-90}}
\caption{Correlation between the equivalent width of [Fe X] 6375~\AA\ and the optical depth of the 
O VII absorption edge.}
\end{figure}
But how can one explain the correlation between the ROSAT spectral index and the equivalent width? 
The ROSAT spectral index might not be a good indicator of the strength of
the warm absorber: no correlation can be found 
between the optical depth of the absorption edges and the ROSAT spectral index
(see Figure~5). 
However, the X-ray spectral shape can be changed if emission and reflection spectra are 
considered (Netzer 1993, 1996), reducing the absorption edges. 
\begin{figure}[htb]
\centerline{\psfig{figure=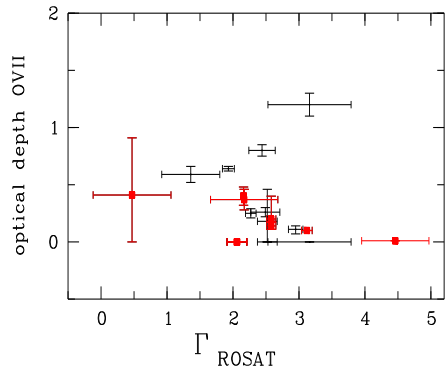,height=2.0truein,width=2.5truein,angle=-90}}
\caption{Correlation between the ROSAT PSPC spectral index and the optical depth of the 
O VII absorption edge.}
\end{figure}

\vspace{-1.0cm}                                                             
    



\begin{thebibliography}{999}
\bibitem{appenzeller88} I. Appenzeller and R. \"Ostreicher {\em AJ} {\bf 95} (1988) 45.
\bibitem{erk} U. Erkens, I. Appenzeller and S. Wagner {\em A\&A} {\bf 323} (1997) 707.
\bibitem{gian} E. Giannuzzo, G.H. Riecke and M.J. Riecke {\em ApJS} {\bf 74} (1990) 371.
\bibitem{nandra} K. Nandra and K.A. Pounds {\em Nature} {\bf 359} (1992) 215.
\bibitem{netzer93} H. Netzer {\em ApJ} {\bf 411} (1993) 594. 
\bibitem{netzer96} H. Netzer {\em ApJ} {\bf 473} (1996) 781. 
\bibitem{osterbrock2} D.E. Osterbrock {\em AJ} {\bf 246} (1981) 696.
\bibitem{puchnarewicz92} E.M. Puchnarewicz et al.\ {\em MNRAS} {\bf 256} (1992) 589. 
\bibitem{reynolds} C.S. Reynolds and A.C. Fabian {\em MNRAS} {\bf 273} (1995) 1167.
\bibitem{reynolds97} C.S. Reynolds {\em MNRAS} {\bf 286} (1997) 513. 
\bibitem{stephens89} S.A. Stephens {\em AJ} {\bf 97} (1989) 10.






\end{thebibliography}
\end{document}